\documentclass[a4paper]{elsarticle}

\usepackage{times}
\usepackage{mathtools}
\usepackage{amsthm}
\usepackage{amssymb}
\usepackage{tikz}
\usepackage{paralist}
\usepackage{fixme,manfnt,mparhack}
\usepackage[basic]{complexity}
\usepackage[T1]{fontenc}
\usepackage{enumerate}
\usepackage{algorithm}
\usepackage{algorithmic}
\usepackage{wasysym}

\usetikzlibrary{positioning,fit,arrows,automata,calc,shapes} 
\tikzset{->,>=stealth',shorten >=1pt,shorten <=1pt,auto,node distance=1cm,
every loop/.style={looseness=6},
initial text={},
every fit/.style={draw,densely dotted,rectangle},
el/.style={font=\footnotesize},
va/.style={font=\scriptsize}}

\newcounter{fixcount}
\newcommand{\defineNote}[3][black!65!green]{\expandafter\def\csname #2\endcsname
##1{\stepcounter{fixcount}\fxwarning{\textcolor{#1}{\textbf{#3}: ##1}}}}
\defineNote{PH}{Paul}
\defineNote{JFR}{Jean-Fran\c cois}

\newtheorem{theorem}{Theorem}

\title{Reachability in succinct one-counter games\tnoteref{t1}}
\tnotetext[t1]{Work supported by ERC inVEST project}
\author[ulb]{Paul Hunter}
\ead{paul.hunter@ulb.ac.be}

\address[ulb]{Universit\'{e} Libre de Bruxelles}

\begin{document}

\begin{abstract}
We consider the reachability problem on transition systems corresponding to succinct one-counter machines, that is, machines where the counter is incremented or decremented by a value given in binary.
\end{abstract}

\maketitle

\section{Preliminaries}
We are interested in reachability problems on transition graphs defined by one-counter machines where:
\begin{itemize}
\item The counter may take any integer value (including negative values);
\item The counter is incremented or decremented by binary weights;
\item Additional transitions are available to the machine if the counter does or does not have value $0$.
\end{itemize}
Previous work~\cite{bjk10,Rei13} has considered other variations on these initial assumptions.

Formally, a transition graph of a one-counter machine (or one-counter graph) is given by a tuple $(V,V_\exists, E, E_0, E_{\neq 0}, q_0, w)$ where
$V$ is a finite set of \emph{states}; $V_\exists \subseteq V$ are the \emph{states of Eve} ($V \setminus V_\exists$ are the \emph{states of Adam}); $E,E_0, E_{\neq 0} \subseteq V \times V$ [$E_0$ ( $E_{\neq 0}$) is the set of edges \emph{(de)activated at $0$}); $q_0 \in V$ is the \emph{initial state}; and $w:E \to \mathbb{Z}$ is the \emph{weight function}.  The (infinite) unweighted arena defined by such a tuple has:
\begin{itemize}
\item Vertex set: $V \times \mathbb{Z}$,
\item Eve's vertices: $V_\exists \times \mathbb{Z}$,
\item Initial vertex: $(q_0,0)$,
\item For every $e=(v,v') \in E$ and $c \in \mathbb{Z}$ an edge from $(v,c)$ to $(v',c+w(e))$, 
\item For every $e=(v,v') \in E_0$ an edge from $(v,0)$ to $(v',0)$, and
\item For every $e=(v,v') \in  E_{\neq 0}$ an edge from $(v,c)$ to $(v',c)$ for $c \neq 0$.
\end{itemize}

\subsection{Reachability problems}
We are interested in the following reachability problems listed in increasing order of difficulty.  They are all known to be in $\EXPSPACE$ and $\EXP\TIME$-hard.  Finite memory strategies suffice for all but parity games.
\paragraph*{Global reachability}
Given $t \in \mathbb{Z}$ does Eve have a strategy to get the counter to value $t$ (in any state)? That is, can she force the play to $(v,t)$ for some $v \in V$? $\EXP\TIME$-hard via straightforward reduction from countdown games.
\paragraph*{Reachability}
Given $t \in \mathbb{Z}$ and $F \subseteq V$ does Eve have a strategy to get the counter to value $t$ in a state of $F$?  
\paragraph*{B\"uchi (repeated reachability)}
Given $F \subseteq V$ does Eve have a strategy to infinitely often have the counter with value $0$ whilst in a state of $F$?
\paragraph*{Parity}
Given a priority function $\Omega:V \to \mathbb{N}$ which defines a priority function on the infinite arena in the obvious way, does Eve win the (infinite) parity game?  Known to be in $\EXPSPACE$ by the result in~\cite{Ser06} which gave a $\PSPACE$ algorithm for parity games on unary-encoded one-counter graphs.

\section{\EXPSPACE-completeness of succinct one-counter games}
We will prove the following:
\begin{theorem}\label{thm:main}
Determining if Eve has a winning strategy in any of these games is \EXPSPACE-complete.
\end{theorem}
From the above results it suffices to show \EXPSPACE-hardness.

\subsection{Simplifying assumptions}
\paragraph*{Activating/Deactivating edges}
It might seem that including edges that are (de)activated when the counter is $0$ might yield a more powerful model, but we can use the antagonistic nature of the game to simulate (de)activating edges.  That is, we activate all transitions but give the other player the ability to punish the (non-)zeroness of the counter.  Figures~\ref{fig:active} and~\ref{fig:active2} show the gadgets that simulate an activating edge $(v,v')$, and Figures~\ref{fig:deactive} and~\ref{fig:deactive2} show gadgets that simulate a deactivating edge $(v,v')$.  In all figures unlabelled edges have weight $0$, square nodes represent states owned by Eve, circle nodes represent states owned by Adam, and all sinks are included in the target set.

\begin{figure}
\begin{minipage}{0.45\textwidth}
\begin{center}
\begin{tikzpicture}[inner sep=2mm, ve/.style={rectangle, draw}, va/.style={circle, draw},vn/.style={regular polygon,regular polygon sides=5,draw}]
\node[ve,draw] (A){$v$};
\node[va, right=of A] (B){};
\node[vn, right=of B] (C){$v'$};
\node[ve,below=0.8 of C,xshift=-5mm] (X){$<$$0$};
\node[ve,above=0.8 of C,xshift=-5mm] (Y){$>$$0$};
\path
 (A) edge  (B)
(B) edge  (C)
(B) edge  (X)
(X) edge[loop, looseness=4, out=-45, in=45] node[el,swap]{-1,$0$}(X)
(B) edge  (Y)
(Y) edge[loop, looseness=4, out=45, in=-45] node[el]{1,$0$}(Y);
\end{tikzpicture}
\end{center}
\caption{Simulating an activating edge from an Eve state}\label{fig:active}
\end{minipage}%
\begin{minipage}{.10\textwidth}
	\phantom{TEXT}
\end{minipage}%
\begin{minipage}{0.45\textwidth}
\begin{center}
\begin{tikzpicture}[inner sep=2mm, ve/.style={rectangle, draw}, va/.style={circle, draw},vn/.style={regular polygon,regular polygon sides=5,draw}]
\node[va,draw] (A){$v$};
\node[ve, right=of A] (B){};
\node[vn, right=of B] (C){$v'$};
\node[ve,below=0.8 of C,xshift=-5mm] (X){$<$$0$};
\node[ve,above=0.8 of C,xshift=-5mm] (Y){$>$$0$};
\path
 (A) edge  (B)
(B) edge  (C)
(B) edge node[el,swap]{$1$}  (X)
(X) edge[loop, looseness=4, out=-45, in=45] node[el,swap]{1,$0$}(X)
(B) edge node[el]{-$1$} (Y)
(Y) edge[loop, looseness=4, out=45, in=-45] node[el]{-1,$0$}(Y);
\end{tikzpicture}
\end{center}
\caption{Simulating an activating edge from an Adam state}\label{fig:active2}
\end{minipage}%
\end{figure}

\begin{figure}
\begin{minipage}{.45\textwidth}
\begin{center}
\begin{tikzpicture}[inner sep=2mm, ve/.style={rectangle, draw}, va/.style={circle, draw},vn/.style={regular polygon,regular polygon sides=5,draw}]
\node[ve] (A){$v$};
\node[va,right=of A] (B){};
\node[ve,above=0.5 of B] (X){$=$$0$};
\node[ve,left=0.5 of X,yshift=10mm] (Y) {$<$$0$};
\node[ve,right=0.5 of X,yshift=10mm] (Z) {$>$$0$};
\node[vn, right=of B] (C){$v'$};
\path
 (A) edge  (B)
(B) edge  (C)
(B) edge (X)
(X) edge node[el]{$1$} (Y)
(X) edge node[el,swap]{-$1$} (Z)
(Y) edge[loop, looseness=4, out=135, in=-135] node[el,swap]{1,0}(Y)
(Z) edge[loop, looseness=4, out=45, in=-45] node[el]{-1,0} (Z);

\end{tikzpicture}
\end{center}
\caption{Simulating a deactivating edge from an Eve state}\label{fig:deactive}
\end{minipage}
\begin{minipage}{.10\textwidth}
	\phantom{TEXT}
\end{minipage}%
\begin{minipage}{.45\textwidth}
\begin{center}
\begin{tikzpicture}[inner sep=2mm, ve/.style={rectangle, draw}, va/.style={circle, draw},vn/.style={regular polygon,regular polygon sides=5,draw}]
\node[va] (A){$v$};
\node[ve,right=of A] (B){};
\node[ve,above=0.8 of B] (X){$=$$0$};
\node[vn, right=of B] (C){$v'$};
\path
 (A) edge  (B)
(B) edge  (C)
(B) edge (X)
(X) edge[loop, looseness=4, out=45, in=135] (X);

\end{tikzpicture}
\end{center}
\caption{Simulating a deactivating edge from an Adam state}\label{fig:deactive2}
\end{minipage}
\end{figure}


\paragraph*{Target set}
We can assume that $F \subseteq V_\exists$ as follows: for every $v \in F \setminus V_\exists$ we add a new vertex $v' \in V_\exists \cap F$ and edge $(v',v)$ [with weight 0] and replace all edges $(u,v)$ with $(u,v')$ [with the same weight].

\subsection{\EXPSPACE-hardness of B\"uchi games}
It is well known that CTL model checking (on a transition system) reduces to a two-player game with a B\"uchi winning condition~\cite{KVW00}. The same reduction shows that CTL model checking on succinct one-counter automata reduces to B\"uchi games on one-counter graphs.  In~\cite{GHOW10}, CTL model checking on succinct one-counter automata was shown to be \EXPSPACE-complete, hence one-counter games with a B\"uchi winning condition are also $\EXPSPACE$-hard.

\subsection{From B\"uchi games to Reachability}

The following lemma, which is readily established using pumping techniques will be useful.

Given a B\"uchi game $G =(V,E,E_0,E_{\neq 0}, q_0,F)$ with target set $F \subseteq V_\exists$, we construct a new one-counter reachability game as follows:
\begin{itemize}
\item The game graph consists of $|F|+1$ copies of $G$ with a $0$-activated edge from $(v,i)$ to $(v,i+1)$ for all $v \in F$ and $1 \leq i \leq |F|$,
\item The initial state is $(q_0,1)$,
\item The target set is $\{(v,|F|+1):v \in F\}$, and
\item The target value is $0$.
\end{itemize}
Clearly Eve wins this game if and only if in the original game she can reach $F$ with counter value $0$ $|F|+1$ times.  Hence if she wins the B\"uchi game she has a winning in the reachability game.  We now show the converse, that is if she can reach $F$ $|F|+1$ times then she can reach some vertex in $F$ with counter value $0$ infinitely often.  More precisely we will show how to defeat any positional (w.r.t. the current state and counter value) strategy for Adam in the original B\"uchi game.  It is well known~\cite{Zie98} that such strategies are sufficient for winning strategies, thus this is sufficient for our result.  Such a strategy has a natural interpretation in the reachability game, so Eve has a counter-strategy to ensure $F$ is visited with counter value $0$ $|F|+1$ times against this strategy.  By the pigeon-hole principle there is some vertex $v \in F$ visited at least twice in the play.  Hence Eve has a strategy (against Adam's strategy) to reach $v$ with counter value $0$ from both $q_0$ and $v$.  Hence Eve can visit $v$ with counter value $0$ infinitely often in the orignal game.

\subsection{From Reachability to Global Reachability}
Given a reachability game $G$ with target set $F \subseteq V_\exists$ and $E_0 = E_{\neq 0} = \emptyset$, we construct a new arena as follows:
\begin{itemize}
\item Double the weights of the edges in $G$;
\item Add a new (initial) vertex $v_0$ and a new sink (with $0$-weighted loops) $v_f$;
\item Add an edge of weight $+1$ from $v_0$ to the original initial vertex;
\item Add edges of weight $-1$ from $F$ to $v_f$.
\end{itemize}
Due to parity arguments the counter can only have value $0$ at $v_f$.  Clearly $v_f$ can be reached with value $0$ if and only if the target set $F$ can be reached with value $0$ in the original game.  Thus Eve wins the Global Reachability game on this new arena if and only if she has a winning strategy in the original Reachability game.
Thus this gives a reduction from Reachability to Global Reachability.  Note that this does not work in the unary case as we utilize the ``long-reach'' ability of doubled weight values to avoid counter values of $0$ in the original game.

\section{Super-exponential counter values}
For one-counter machines without alternation (i.e.\ one player games) it is known~\cite[Lemma 42]{LLT04} that the reachability problem can be solved without the counter value exceeding an exponential bound.  Our results show that such a bound in the case of alternating machines is unlikely -- it would yield an alternating \PSPACE (i.e.\ \EXP\TIME) algorithm, thereby implying $\EXP\TIME = \EXPSPACE$.  We now give a concrete example that shows super-exponential counter values are in fact necessary in succinct one-counter games.

\paragraph*{Game summary}
The game $G_n$ proceeds as follows:
\begin{enumerate}
\item Eve increments the counter to a multiple of $2^n$, $M\cdot 2^n$,
\item Adam chooses some odd $m \in (0,2^n)$ and adds it to the counter,
\item Eve removes a multiple (at least one) of $m \cdot 2^n$ from the counter, and
\item Eve removes some $m' \in (0,2^n)$ from the counter.
\end{enumerate}

\paragraph*{Implementation}
Step 1 is implemented by a single Eve vertex with a loop of weight $2^n$.  Steps 2 and 4 can be implemented by a sequence of $n$ nodes (belonging to the relevant player) where two edges from the $i$-th to $(i+1)$-th vertex allow the relevant player to choose the $i$-th bit.  Step 3 is implemented by a sub-game of $n$ rounds repeated as often as Eve chooses (but at least once).  In the $i$-th round of the subgame Adam chooses the $i$-th bit $b$ of $m$.  If he chooses correctly then $b \cdot 2^{i+n}$ is subtracted from the counter and the subgame continues, if he chooses incorrectly then the $i$-th bit is cleared, Eve exits the subgame and clears all but the $i$-th bit.  Note that if Eve tries to exit when Adam chose correctly then the $i$-th bit is never cleared so Eve is unable to reach a counter value of $0$.

\paragraph*{Correctness}
Clearly Eve can reach a counter value of $0$ at the end if and only if $m|M$,  $M>0$ and $m'=m$.  Thus in order to win the game $M$ must be a non-zero multiple of all odd numbers in $(0,2^n)$, in particular it is at least the product of all (odd) primes less than $2^n$.  Hence $M \geq 2^{\pi(2^n)-1}$, where $\pi(x)$ is the number of primes less than $x$.  Using the standard lower bound of $\frac{x}{\ln x}$ for $\pi(x)$~\cite{rs62}, we have $M \geq 2^{2^n/n-1}$, and hence the counter necessarily attains super-exponential values.


\end{document}